\documentstyle[multicol,aps,epsf,amsmath]{revtex}
\begin{document}
\draft
\def\av#1{\langle#1\rangle}
\def\a{\alpha}
\def\etal{{\it et al.}}
\def\pc{p_{\rm c}}
\def\df{d_{\rm f}}
\def\K{{\tilde K}}
\def\p{{\tilde p}}
\def\P{{\tilde P}}
\title {Breakdown of the Internet under intentional attack}
\author{Reuven~Cohen$^{1}$
\footnote  {{\bf e-mail:} cohenr@shoshi.ph.biu.ac.il}, Keren~Erez$^1$,
Daniel~ben-Avraham$^2$, and Shlomo Havlin$^1$}
\address{$^1$Minerva Center and Department of Physics, Bar-Ilan university,
Ramat-Gan, Israel}
\address{$^2$Department of Physics,Clarkson University,
Potsdam NY 13699-5820,
USA}
\maketitle
\begin{abstract}
We study the tolerance of random networks to intentional attack,
whereby a fraction $p$ of the most connected sites is removed.   We focus
on scale-free networks, having connectivity distribution $P(k)\sim k^{-\a}$
(where $k$ is the site connectivity),
and use percolation theory to study analytically and numerically the
critical fraction $\pc$ needed for the disintegration of the network, as
well as the size of the largest connected cluster.  We find that even
networks with $\a\leq 3$, known to be resilient to random removal of
sites, are sensitive to intentional attack.  We also argue that, near
criticality, the average distance between sites in the spanning (largest)
cluster scales with its mass, $M$, as $\sqrt{M}$, rather than as
$\log_k M$, as expected for random networks away from criticality. Thus,
the disruptive effects of intentional attack become relevant even
before the critical threshold is reached.
\end{abstract}
\begin{multicols}{2}
The question of stability of scale-free random networks to removal of a
fraction of their sites, especially in the context of the Internet, has
been recently of interest~\cite{bar2,cohen,cal}.   The Internet can be
viewed as a special case of a random, scale-free network, where the
probability of a site to be connected to $k$ other sites follows a
power-law: $P(k)\sim k^{-\a}$ ($\a\approx2.5$, for the Internet).
It is now well established that if a fraction $p$ of the
sites is removed {\em randomly}, then for $\a>3$ there exists a
critical threshold, $\pc$, such that for $p>\pc$ the network
disintegrates; networks with $\a\leq3$ are more resilient and do not
undergo this transition, although finite networks (such as the
Internet) may be eventually disrupted when nearly all of
their sites are removed, as shown numerically in~\cite{bar2,cohen}, and
analytically in~\cite{cohen}.

Albert \etal,~\cite{bar2} have introduced a model for intentional attack,
or sabotage of random networks: the removal of sites is not random, but
rather sites with the highest connectivity are targeted first.  Their
numerical simulations suggest that scale-free networks are highly
sensitive to this kind of attack. In this Letter we study the
problem of intentional attack in scale-free networks.  Our study
focuses on the exact value of the critical fraction needed for
disruption and the size of the remaining largest connected cluster.
We also study the distance between sites on this cluster near the
transition. We find, both analytically and numerically, that
scale-free networks are highly sensitive to sabotage of a small fraction
of the sites, for all values of $\a$, lending
support to the view of Albert \etal,~\cite{bar2}.

In a recent paper~\cite{cohen} we have studied the properties of the
percolation phase transition in scale-free random networks, and
applied a general criterion for the existence of a spanning cluster
(a cluster whose size is proportional to the size of the
network)~\cite{molloy,cohen}:
\begin{equation}
\label{kappa}
\kappa\equiv {\av{k^2}\over\av{k}}=2\;.
\end {equation}
Here $k$ is the site connectivity, and averages, indicated by angular
brackets, are taken over all sites of the network.   When a fraction $p$
of the sites are randomly removed (or a fraction $p$ of the
links are removed, or lead to deleted sites), the distribution of site
connectivity is changed from the original $P(k)$ to a new distribution
$\P(k)$:
\begin{equation}
\label{new_pk}
\P(k)=\sum_{{k_0}\geq k}^{K} P(k_0)\binom{k_0}{k}(1-p)^k p^{{k_0}-k}\;.
\end {equation}
Using this criterion together with Eq.~(\ref{kappa}), the critical
threshold
$p=\pc$ is found to be:
\begin{equation}
\label{perc}
1-\pc={1 \over \kappa_0-1}\;,
\end {equation}
where $\kappa_0=\av{k_0^2}/\av{k_0}$ is calculated from the original
connectivity  distribution, before the removal of any sites~\cite{cohen}.

A wide range of networks, including the Internet, have site connectivities
which follow a power-law distribution~\cite{bar2,fal,remark}:
\begin{equation}
\label {SF1}
P(k)=ck^{-\a}, \quad k=m,m+1,...,K,
\end {equation}
where $k=m$ is the minimal connectivity and $k=K$ is an effective
connectivity cutoff present
in finite networks.  For the  distribution~(\ref{SF1}),
$\kappa_0$ can be approximated by~\cite{remark}:
\begin{equation}
\label{SF_perc}
\kappa_0 = \biggl({2-\a\over 3-\a}\biggr)
{{K^{3-\a}-m^{3-\a}}\over{K^{2-\a}-m^{2-\a}}}\;.
\end{equation}
This, together with Eq.~(2), was used to show that networks with
$\a\leq 3$, which  have a divergent second moment, are resilient to random
deletion of sites~\cite{cohen}. Indeed, when the number of sites in such
networks $N\to\infty$, then the upper cutoff $K\to\infty$, and there
exists a spanning cluster for all values of $p<1$. Another approach,
based on generating functions, was introduced in~\cite{newman} and was
used to study a similar problem in~\cite{cal}.

Consider now intentional attack, or sabotage~\cite{bar2}, whereby a
fraction  $p$ of the sites with the highest connectivity is removed.  (The
links emanating from the sites are removed as well.) This has the
following effect: (a) the cutoff connectivity
$K$ reduces to some new value, $\K<K$, and (b) the connectivity
distribution of the remaining sites is no longer scale-free, but is
changed, because of the removal of many of their links.  The upper
cutoff $K$ before the attack may be estimated from
\begin{equation}
\sum_{k=K}^\infty P(k)={1\over N}\;,
\end {equation}
where $N$ is the total number of sites in the network.
Similarly, the new cutoff $\K$, after the attack, can be estimated from
\begin{equation}
\label{cutoff}
\sum_{k=\K}^K P(k)=\sum_{k=\K}^{\infty}P(k)-\frac{1}{N}=p\;.
\end {equation}
If the size of the system is large, $N\gg1/p$, the original cutoff $K$ may
be safely ignored.  We can then obtain $\K$ approximately by replacing
the sum with an integral~\cite{remark}:
\begin{equation}
\label {attack_K}
\K=mp^{1/(1-\a)}\;.
\end {equation}

We estimate the impact of the attack on the distribution of the
remaining sites as follows.  The removal of a fraction $p$ of the sites
with the highest connectivity results in a random removal of links from
the remaining sites --- links that had connected the removed sites with
the remaining sites.  The probability $\p$ of a link leading to a deleted
site equals the ratio of the number of links belonging to deleted sites
to the total number of links:
\begin{equation}
\label {attack_p}
\p=\sum_{k=\K}^K {kP(k)\over\av{k_0}},
\end {equation}
where $\av{k_0}$ is the initial average connectivity. With the usual
continuous approximation, and neglecting $K$, this yields
\begin{equation}
\label {attack_p2}
\p=\biggl(\frac{\K}{m}\biggr)^{2-\a}={p^{(2-\a)/(1-\a)}}\;,
\end {equation}
for $\a>2$. For $\a=2$, $\p\to 1$, since just a few nodes
of very high connectivity control the entire connectedness of
the system.  Indeed, consider a finite system of $N$ sites and $\a=2$.
The upper cutoff $K\approx N$ must then be taken into account, and
approximating Eq.~(\ref{attack_p}) by an integral yields
$\p=\ln(Np/m)$.  That is, for $\a=2$,  very small
values of $p$ are needed to destroy an arbitrarily large fraction of the
links as $N\to\infty$.

With these results known we can compute the effect of
intentional attack, using the theory previously developed
for random removal of sites~\cite{cohen}.  Essentially, the network after
attack is equivalent to a scale-free network with cutoff $\K$,
that has undergone random removal of a fraction $\p$ of its sites.
This can be seen as the result of two processes: (a)~Removal of the
highest connectivity sites reduces the upper cutoff.  Since this effect
changes the connectivity distribution, $\kappa_0$ needs to be
recalculated accordingly.  (b)~Removal of the links
leading to the removed sites. The probability of removing a link is
$\p$ --- the probability of a randomly chosen link to lead to one of the
removed sites --- and all links have the same probability of being
deleted. Since this effect has the influence on the probability distribution
described in Eq. (\ref{new_pk}), the result in Eq. (\ref{perc}) can be used,
with $\p$ replacing $p$. (Notice that for random site deletion the probability
of a link leading to a deleted site is identical to the fraction of deleted
sites.)

Although the number of nodes removed in intentional attack is different
than in the random breakdown model, this affects the size of the spanning
cluster (see below) but not the
critical point.   This is because the transition point is defined as
the point where the spanning cluster becomes a finite fraction of the
whole network.  A finite fraction of the remaining nodes is
also a finite fraction of the original network, so the difference has no
effect on $\pc$.

\begin{figure}
\narrowtext
\epsfxsize=2.5in
\hskip 0.5in
\epsfbox{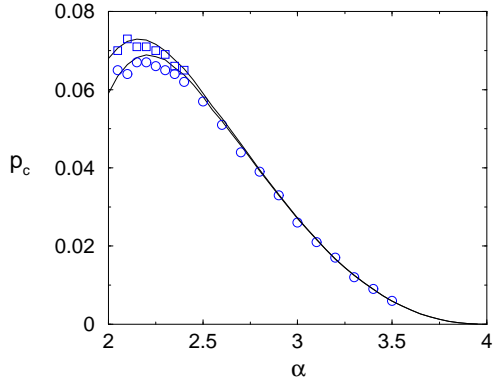}
\vskip 0.15in
\caption{Critical probability, $\pc$, as a function of $\a$, for
networks of size $N=500,000$ (circles) and $N=64,000$ (squares). Lines
represent the analytical solution, obtained from Eqs.~(\ref{cutoff}) and
(\ref{attack_p}).
\label{pc} }
\end{figure}

We therefore use Eqs.~(\ref{perc}) and (\ref{SF_perc}), but with
$\p=(\K/m)^{2-\a}$ and $\K$ replacing $\pc$ and $K$.  This yields the
equation:
\begin{equation}
(\K/m)^{2-\a}-2=  \case{2-\a}{3-\a}m[(\K/m)^{3-\a}-1]\;,
\end{equation}
which can be solved numerically to obtain $\K(m,\a)$, and then
$\pc(m,\a)$ can be retrieved from
Eq.~(\ref{attack_K}).  In Fig.~\ref{pc} we plot $\pc$ --- the critical
fraction of sites needed to be removed in the sabotage strategy to
disrupt the network --- computed in this fashion, and compared to
results from numerical simulations.  A phase transition exists (at a
finite $\pc$) for all $\a>2$.  The decline in $\pc$ for large $\a$ is
explained from the fact that as $\a$ increases the spanning cluster
becomes smaller in size, even before attack.  (Furthermore, for
$m<2$ the original network is disconnected for some large enough $\a$.)
The decline in $\pc$ as $\a\to 2$ results from the critically high
connectivity of just a few sites: their removal disrupts the whole
network.  This was already argued in~\cite{bar2}.  We note that for
infinite systems $\pc\to 0$ as $\a\to 2$.  The critical fraction $\pc$
is rather sensitive to the lower connectivity cutoff $m$.  For larger $m$
(the case of $m=1$ is shown in Fig.~\ref{pc}) the networks are more
robust, though they still undergo a transition at a finite $\pc$.

\begin{figure}
\narrowtext
\epsfxsize=2.5in
\hskip 0.5in
\epsfbox{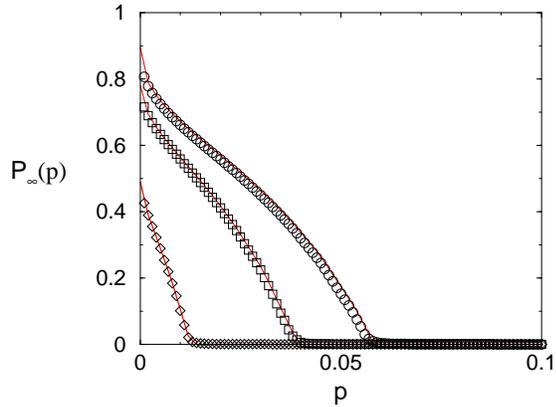}
\vskip 0.15in
\caption{Fraction of sites belonging to the spanning cluster, $P_\infty$,
as a function of the fraction of removed sites, $p$,
for networks with
$\a=2.5$ (circles), $\a=2.8$ (squares), and $\a=3.3$ (diamonds). Lines
represent the  analytical result from
Eqs.~(\ref{p_infty}) and (\ref{u}).   Both the simulation and analysis are
for system size $N=500,000$.
\label{p_inf}}
\end{figure}

The size of the spanning cluster as a fraction of the number of
undeleted sites, $P_{\infty}(p)$, can be calculated using the methods
introduced in~\cite{molloy2} and developed in~\cite{cal,newman}.
Following closely the derivation in~\cite{newman}, a generating function
is built for the connectivity  distribution:
\begin{equation}
G_0 (x)=\sum_{k=0}^\infty P(k) x^k.
\end {equation}
The probability of reaching a site with connectivity $k$ by following a
specific link is $k P(k)/\av{k}$~\cite{cohen,cal,molloy,newman}, and the
corresponding generating function is
\begin{equation}
G_1(x)={{\sum k P(k) x^{k-1}}\over {\sum k P(k)}} =
\frac{d}{dx}G_0(x)/\av{k}\;.
\end {equation}
Hence, the generating function for the probability of reaching a
branch of  a given size by following a link is
\begin{equation}
H_1 (x)=x G_1 (H_1 (x))\;,
\end {equation}
while the generating function for the size of a component is
\begin{equation}
H_0 (x)=x G_0 (H_1 (x))\;.
\end {equation}
Then, $P_\infty=1-H_0(1)$, since
$H_0$ contains only the finite-size clusters.   It follows that
\begin{equation}
\label {p_infty}
P_\infty (p)=1-\sum_{k=0}^{\infty} \P(k) u^k\;,
\end {equation}
where $u\equiv H_1(1)$ is the smallest positive root (found numerically) of
\begin{equation}
\label{u}
\av{k}u=\sum_{k=0}^{\infty}kP(k)u^{k-1}\;.
\end{equation}
We argue that the same holds true after attack, but the sums in
Eqs.~(\ref{p_infty}) and (\ref{u}) should run only up to $k=\K$, and the
original distribution $P(k)$ should be replaced by the new connectivity
distribution~\cite{cohen}:
\begin{equation}
\label {SF_perc_dist}
\P(k)=\sum_{{k_0}=m}^{\K} P(k_0)\binom{k_0}{k}(1-\p)^k \p^{{k_0}-k}\;.
\end {equation}
The actual fraction of removed sites, $p$,  is inconsequential since we
calculate the size of the infinite cluster relative to  the number of {\em
undeleted} sites.
$P_\infty$ evaluated in this fashion, and compared to numerical
simulations, is shown in Fig.~\ref{p_inf}.

Since intentional attack leads always to a finite cutoff which does not
scale with the system size, all the moments are finite and
Eqs.~(\ref{p_infty}) and (\ref{u}) are well behaved. Therefore, there
always exists a linear term in the series expansion of $P_\infty(p)$.
Hence, near the critical point
$P_\infty \sim |\pc-p|^\beta$, where $\beta=1$~\cite{remark2}.

\begin{figure}
\narrowtext
\epsfxsize=2.5in
\hskip 0.5in
\epsfbox{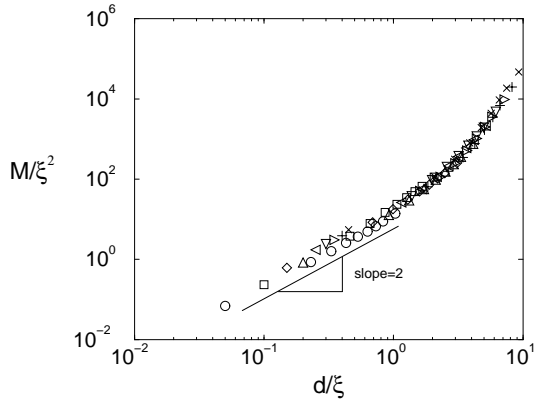}
\vskip 0.15in
\caption{Mass (number of sites), $M$, as a function of distance, $d$,
on the spanning cluster. The correlation length is $\xi=|p-p_c|^{-1}$.
Note that for $d/\xi<1$, the slope is $2$, corresponding to the behavior
in the critical regime, while for $d/\xi>1$, $M$ grows exponentially with
$d$, corresponding to the well connected regime.
\label{mr}}
\end{figure}

Finally, we consider the distance between sites in the spanning cluster.
The behavior of this quantity in diluted networks is different from
that at the highly connected regime. The average distance between
two sites in the spanning  cluster of a highly connected network is
proportional to $\log_k N$, where $k$ is the average connectivity and
$N$ is  the number of sites~\cite{bal}. This has also been shown to hold
for scale-free and general networks~\cite{newman}. However, the diluted
case is essentially the  same as infinite-dimensional percolation. In
this case, there is no notion of geometrical distance (since the graph is
not embedded in an Euclidean space), but only of a distance along the
graph (which is the  shortest distance along bonds). It is known from
infinite-dimensional  percolation theory that the fractal dimension at
criticality is
$\df =2$~\cite{havlin}.
Therefore the average (chemical) distance $d$ between pairs of sites on the
spanning cluster at criticality behaves as
\begin {equation}
d \sim \sqrt{M}\;,
\end{equation}
where $M$ is the number of sites in the spanning cluster. This is analogous
to percolation in finite dimensions, where in lengthscales smaller than
the correlation length the cluster is a fractal with dimension $\df$ and
above the correlation length the cluster is homogeneous and has the
dimension of the embedding space. In our infinite-dimensional case, the
crossover between these two behaviors occurs around the correlation
length $\xi\approx|\pc-p|$, as can be seen in  Fig.~\ref{mr}.

In summary, we have shown that scale-free networks are highly sensitive to
intentional attack.  This is true even for networks with $\a<3$,
which are known to be resilient to random removal of their sites.
The high sensitivity near
$\a=2$ results from the presence of just a few sites with connectivity
comparable to the size of the system:  their removal disrupts the whole
network.  We note that while the cutoff $\K$ must reach a typically small
number before the network is disrupted, this is achieved
with a modest removal fraction $p$ (Eq.~\ref{attack_K}).
The effect of sabotage on the connectivity distribution of the remaining sites
after the attack and thus the relation between $\p$, $\K$, and $p$ (Eqs.
(\ref{cutoff}) and (\ref{attack_p}) ) is found explicitly in our approach.
This effect is particularly important near the borderline case of $\a=2$. We
have also shown that the average distance between pairs of sites in the
spanning cluster  grows dramatically near criticality.  This makes
communication very inefficient, even before the spanning cluster is
completely disrupted.

\acknowledgments
We thank the National Science Foundation for support, under grant
PHY-9820569  (D.b.-A.).

\end{multicols}

\begin{thebibliography}{99}
\bibitem{remark2}
   However, for random breakdown when the upper cutoff $K\rightarrow\infty$,
   there are diverging moments. Using Abelian theories for calculating the
   sums in Eqs. (\ref{p_infty}) and (\ref{u}), it can be shown that
   for $3<\alpha<4$, $\beta=1/(\alpha-3)$ while for $\alpha>4$, $\beta=1$ as
   in regular infinite-dimensional percolation.
\end{thebibliography}
\end{document}